# Big and Small

**R D Ekers**[1]
*CSIRO-ATNF*
*Sydney, NSW, Australia*
*E-mail: ron.ekers@csiro.au*

**Abstract**

Technology leads discovery in astronomy, as in all other areas of science, so growth in technology leads to the continual stream of new discoveries which makes our field so fascinating. Derek de Solla Price had analysed the discovery process in science in the 1960s and he introduced the terms 'Little Science' and 'Big Science' as part of his discussion of the role of exponential growth in science. I will show how the development of astronomical facilities has followed this same trend from 'Little Science' to 'Big Science' as a field matures. We can see this in the discoveries resulting in Nobel Prizes in astronomy. A more detailed analysis of discoveries in radio astronomy shows the same effect. I include a digression to look at how science progresses, comparing the roles of prediction, serendipity, measurement and explanation. Finally I comment on the differences between the 'Big Science' culture in Physics and in Astronomy.

[1] Speaker

# 1. Exponential Growth in Science

Harwit [1] showed that most important discoveries in astronomy result from technical innovation. The discoveries peak soon after new technology appears, and usually within 5 years of the technical capability. Instruments used for discoveries are often built by the observer. He also noted that new astronomical phenomena are more frequently found by researchers trained outside astronomy.

It had already been well established that most scientific advances follow technical innovation in other areas of science. In 1960 de Solla Price [2] applied quantitative measurement to the progress of science (scientometrics) and reached the conclusion that most scientific advances follow laboratory experiments. His analysis also showed that the normal mode of growth of science is exponential. Derek de Solla Price had worked as a teacher of applied mathematics at Raffles College (University of Singapore) in 1948 and it was there that he formulated his theory on the exponential growth of science [2]. The idea occurred to him when he noticed the exponential growth in stacks of the complete set Philosophical Transactions of the Royal Society between 1665 and 1850, which he had in his home while Raffles College had its library built [3]. Historical examples of exponential growth included the rate of discovery of elements and the number of universities founded in Europe. Some more recent examples of exponential growth and their doubling times are: power consumption (10 years), overseas telephone calls (5 years), particle accelerator beam energy (2 years) and internet hosts (1 year). These are all much faster than the underlying growth rates such as population (50 years), GNP (20 years).

Such exponential growth cannot continue indefinitely and when it reaches a ceiling de Solla Price [4] noted three possible consequences:

1. Progress in this area of development becomes chaotic,

2. The area of research dies out,

3. There is a reorganization or change in technology which results in a new period of exponential growth and research flourishes.

A rather simplified conclusion to draw from this is that any field which has not maintained an exponential growth has now died out, so current active research areas are all still in an exponential growth phase. Furthermore, to maintain the exponential the continual introduction of new technology is required since just refining existing technology plateaus out.

## 1.1 Livingstone Curve

A famous example which illustrates this very well is the rate of increase of operating energy in particle accelerators by Livingston and Blewett [5]. Starting in 1930, each particle accelerator technology provided exponential growth up to a ceiling when a new technology was introduced. The envelope of the set of curves is itself an exponential with an increase in energy of $10^{10}$ in 60 years. This has been recently updated by Riesselmann to include the Large Hadron Collider [6]. This example of exponential growth, originally presented by Fermi in 1954, has become known as the 'Livingstone Curve'.

## 1.2 Moore's Law

To this we can add the now famous 'Moore's Law' for computing devices (more precisely for transistors on a chip). In 1965 Gordon Moore (co-founder of Intel) noted that the transistor density of semiconductor chips doubled roughly every 1-2 years [7], Figure 1, later refined to doubling every 18 months and this exponential growth has been maintained for the last 40 years [8].

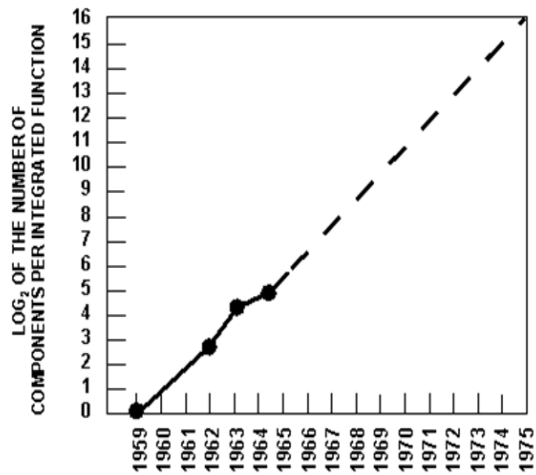

**Figure 1**: Microprocessor performance, original Moore's Law plot [7]

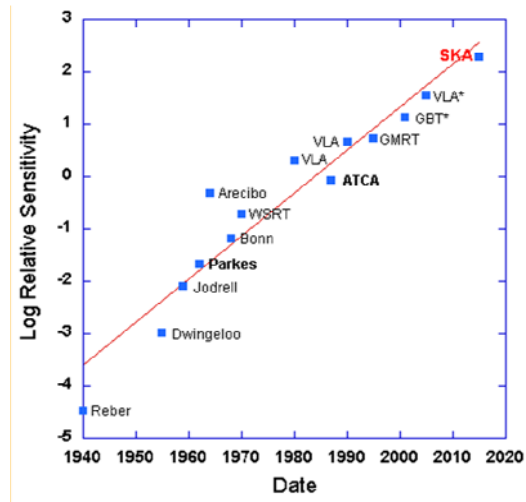

**Figure 2**: Radio Telescope Sensitivity vs. time. Points are the relative continuum sensitivity when the telescopes were built, or after major upgrades. VLA* is the EVLA upgrade.
 SKA is the proposed sensitivity for a telescope which has not yet been built.

## 1.3 Radio telescopes

Figure 2 plots the sensitivity of telescopes used for radio astronomy since the discovery of extra-terrestrial radio emission in 1940. It has been exponential with an increase in sensitivity of $10^5$ since 1940, doubling every three years. Also in this case we can see particular radio telescope technologies reaching ceilings and new technologies being introduced e.g., the transition from huge single dishes to arrays of smaller dishes in the 1980s.

## 1.4 How to Maintain Exponential Growth?

If the improvement in sensitivity has reached a ceiling the rates of new discoveries will decline and the field will become uninteresting and die out. On the other hand, if we can shift to new technology or find new ways to organize our resources the exponential increase in sensitivity can continue. Do we have such new technology to continue the exponential improvement? In radioastronomy the combination of transistor amplifiers and their large scale integration into complex systems which can be duplicated inexpensively provides one of the keys for change. The other key technology is the computing capacity to apply digital processing at high bandwidth thereby realizing processes such as multiple adaptive beam formation and active interference rejection in ways not previously conceivable. Finally, the move to international facilities such as the proposed SKA will also be needed to avoid the resource ceiling.

## 2. From *Little* Science to *Big* Science:

Exponential growth cannot continue indefinitely without hitting the ceiling on cost or available resources to continue. As discussed in the previous section, sometimes technical innovation comes to the rescue but de Solla Price also recognized the important role played by the transition from Individual Researcher to Institute to National Facility and, finally, to International Facility, each step removing a resource limitation ceiling. He coined the terms 'little science' and 'big science' to describe the two extremes.

- Institutional Facilities are built to enable research on a scale which no individual can afford.
- National Facilities are built to enable research on a scale which no single institute can afford.
- International Facilities are built to enable research on a scale which no single nation can afford.

While the progression clearly involves an increasing big science component, this doesn't mean the small science role has to disappear, and as discussed in the following sections, it is important that it doesn't disappear. Provided a field remains active there will be a rich diversity of the scale of the facilities.

### 2.1 Big Science

In addition to the obvious resource advantage, the big national and international facilities have other advantages. The global linkages broaden our knowledge base and provide cross fertilisation between diverse communities. Networking can now provide access to the facility for a wide community of users and these users from different backgrounds will also interact with each other. The development of international facilities is an excellent way for all of us to learn to play together.

In addition to the direct scientific advantages of international facilities we have important indirect advantages from the global collaboration. In most nations government funding will be linked, either directly or indirectly, to wealth creation through industry involvement. Large astronomy facilities can achieve this in a number of ways:
- Industries or nations can use the open access facilities to showcase technology
- International involvement means the technology used will be benchmarked against international standards
- Astronomers are seen as sophisticated end users who can provide valuable feed-back and incentives for the technology development

- Industry links will be developed between organisations in a non-competitive environment
- Technology transfer will stimulate innovation

**2.2 The Continuing Case for Small Science**

However it is not all good news - big science facilities are expensive so they need to be common-user facilities to justify their cost. Smaller, specialized instruments which are more cost effective will only be useful for special projects and cannot support a large community of users. In comparison, small projects can be agile; small teams can move fast and adapt to changing circumstances. Multiple small projects will create more educational opportunities and will be essential for the next generation of big facility designers. Big projects necessarily have large bureaucracies which tend to crush creative entrepreneurship, innovative ideas and innovative individuals. It is also less likely that membership in a big team will be as rewarding to all individuals, although for others the team environment can be very satisfying.

**2.3 Small Science on Big Telescopes**

Compromise is possible and if we plan carefully we can still use big telescopes to do small science. In fact this is one of the distinguishing characteristics of the astronomy culture compared to the particle physics culture where big facilities are used by huge teams focussed on a small number of problems. Simon White has triggered a vigorous ongoing debate on this topic [6].

Multiple small teams or individuals can use common-user facilities and small groups can build instruments and develop specialised software packages and processing techniques for big facilities. This requires appropriate management structures and funding support to maintain the small groups.

# 3. Evidence for the Impact of Instrumental Development on Advances in Astronomy

**3.1 Antikythera Machine**

Horace Walpole coined the word serendipity to describe the discovery of a book or information that you were not seeking. While looking into de Solla Price's scientometrics I discovered his early research on the Antikythera Mechanism. A century ago, pieces of a strange mechanism with bronze gears and dials were recovered from an ancient shipwreck on the island of Antikythera off the coast of Greece. Historians of science concluded that this was an instrument that, originating in 80 B.C, calculated and illustrated astronomical information, particularly phases of the Moon, planetary motions and even predicted eclipses [3]. While this might not quite be classified as big science it was an extraordinary technology development for the time. This technology disappeared for a millennium, a sobering reminder that our technology can also go backwards and that exponential growth is not guaranteed.

### 3.2 Nobel Prize Discoveries

Table 1 lists the 10 astronomical discoveries which have resulted in Nobel prizes. In Figure 3 I have plotted these discoveries against the discovery date and a subjective indication of the relative scale of the instrument or research group involved. It is quite clear from Figure 3 that the role of Big Science facilities in making discoveries increases in importance with time.

**Table1**: Nobel Prizes for astronomy

| Prize | Experiment | Subject | Laureates |
|---|---|---|---|
| 1936 | 1912 | Cosmic Rays | Victor Franz Hess (shared) |
| 1974 | 1960 | Aperture Synthesis | Sir Martin Ryle |
| 1974 | 1967 | Pulsars | Antony Hewish |
| 1978 | 1965 | CMB | Arno A. Penzias, Robert W. Wilson |
| 1983 | 1931 | Stellar Evolution | Subrahmanyan Chandrasekhar |
| 1983 | 1950 | Chemical Elements | William Alfred Fowler |
| 1993 | 1974-78 | Gravitational Radiation | Russell A. Hulse, Joseph H. Taylor, Jr. |
| 2002 | 1987 | Cosmic Neutrinos | Raymond Davis, Jr., Masatoshi Koshiba |
| 2002 | 1962,70 | Cosmic X-rays | Riccardo Giacconi |
| 2006 | 1989 | CMB | John C. Mather, George F. Smoot |

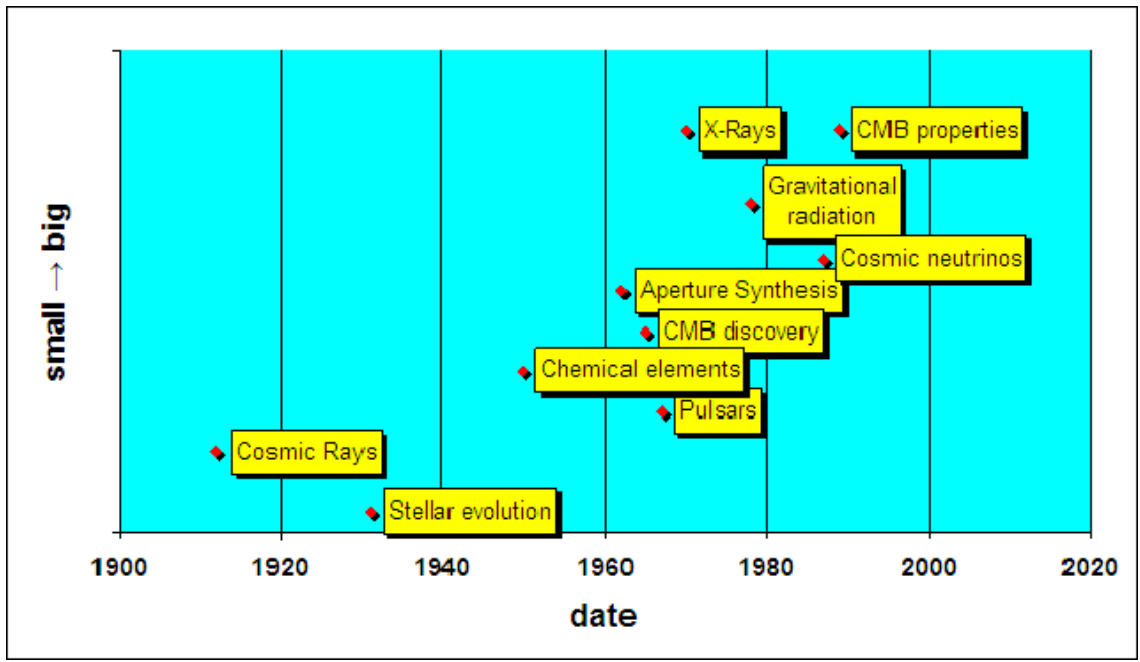

**Figure 3:** Nobel Prizes in astronomy vs. date of discovery with an indication of the relative scale of the experiment

### 3.3 Discovery of Cosmic Rays

Cosmic ray research began in 1912 when Victor Hess, of the Vienna University, flew in a balloon with his electroscope to an altitude of about 16,000 ft. He discovered evidence of a very penetrating radiation (cosmic rays) coming from outside our atmosphere. In 1936, Hess was awarded the Nobel Prize for this discovery. It was clearly a small science experiment. The field has prospered with sporadic bursts of activity since then but is now very much alive with the creation of the international big science facilities, the Pierre Auger Observatory built in Argentina and its northern hemisphere counterpart being built in Utah to search for the highest energy cosmic rays..

### 3.4 Discovery of the Cosmic Microwave Background, CMB

As discussed by Kellermann et al [11], the discovery of the CMB was a serendipitous observation of a predicted phenomenon which resulted in the award of the 1978 Nobel Prize to Penzias and Wilson for their 1965 discovery of the Big Bang radiation with the Bell Telephone Laboratory horn. In 1989 the COBE satellite measured properties of the microwave background and the 2006 Nobel Prize was awarded to John Mather for the spectrum and to George Smoot for the anisotropy of the CMB. The initial discovery was made by a small team using a modest but very state-of-the-art telescope at the Bell Telephone Laboratory. The follow-up observation was made with a NASA funded space telescope built by a large team and had clearly entered the 'Big Science' era.

### 3.5 Pulsars and Gravitational Radiation

The initial discovery of pulsars by Hewish and Bell in 1968 was made with a modest (institute scale) telescope but the discovery of the binary pulsar by Hulse and Taylor in 1974 and its use to detect gravitational radiation in 1978 required the Arecibo telescope which is operated as a US national facility and is the largest aperture telescope ever built.

## 4. Discoveries in Radio Astronomy

The beginning of radio astronomy provides excellent examples of discoveries made by exploring the unknown [11]. Wilkinson et al [12] included a tabulation of the key discoveries in radio astronomy since the beginning of the field in 1933 to 2000. Figure 4 (a) plots these discoveries against time, comparing the discoveries made with special purpose instruments with those made on the larger general user facilities. It is clear that the number of discoveries made with special purpose instruments has declined with time. Figure 4 (b) shows that serendipitous discoveries are more prevalent at the inception of a new branch of science.

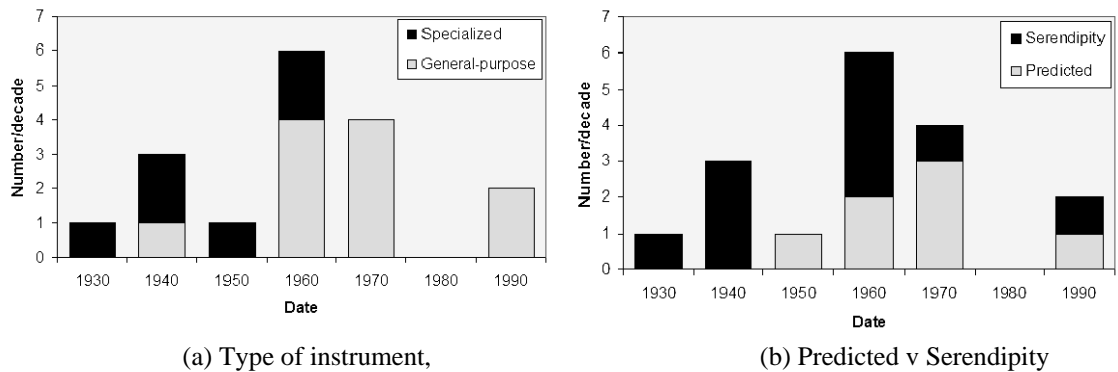

(a) Type of instrument,　　　　　　　　(b) Predicted v Serendipity

**Figure 4**: Key Discoveries in Radio Astronomy from [12]

## 5. The Analysis vs. Discovery dilemma

The preceding examples focus on the discovery of new phenomena but most astronomical research involves the analysis of known phenomena. Can we optimise our telescopes to do both? We have a similar dilemma when we look at the criteria used to design new telescopes, do we base designs on the known phenomena or do we design to maintain the flow of new discoveries? [11].

### 5.1 Analysis of Known Phenomena

Measurements are made to understand the way a known class of object works. For these the research involves explaining and measuring. This requires stable observational tools and progresses by incremental steps. Common user facilities are mostly used as analytic tools. In this process discoveries of the unexpected may still be made but good understanding of the instrument is important to separate unexpected real events from instrumental errors which are the most common cause of an unusual result.

### 5.2 Discovery of New Phenomena

New phenomena (either predicted, or unanticipated) are found. This requires new observational approaches, and expanded parameter space.

### 5.2.1 Prediction or Serendipity?

There are predicted new phenomena which are either confirmed by an observation or are observed accidentally but still confirm an existing prediction. There are also serendipitous discoveries of the unexpected which lead to new and expanded understanding. These are often the trigger for a new theory, or the resurrection of an old one. Sometimes predictions are triggers for other discoveries or lead to technology developments which open up other observational opportunities.

The 21cm Hydrogen Line was predicted by van der Hulst in 1944 and detected by Ewen & Purcell at Harvard in 1951. The gravitational radiation predicted by Einstein was observed by Hulse and Taylor using a binary pulsar which they had discovered in 1974 and for which they were awarded a Nobel prize in 1993.

### 5.2.2 Predicting v Explaining

When there are only a few degrees of freedom in a theory there will be only a few solutions. Predictions are then possible and any observational constraint will be important, e.g. The CMB theory is constrained by a relatively small number of observations and is making testable predictions about the CMB structure.

When there are many degrees of freedom in a complex phenomenon there are many solutions so that many predictions are possible, consequently they are less valuable. In this case the observations will generally guide the interpretation.

### 5.3 Managing Scientific Research

Irving Langmuir, General Electric, (& New Mexico Tech) in the 1950s noted that you can't plan to make discoveries but you can plan a work environment which increases the chance of discovery. He argued that you need to encourage individual freedom to explore, provide opportunities for discussion in an open environment and encourage contacts outside the field. He also argued that it is necessary to avoid the over protection of information, over management, and lack of time to pursue other ideas.

## 6. Big Science Culture: Physics vs. Astronomy

We have a vigorous ongoing debate around the big science physics culture and whether it will have a good or a bad influence on the astronomy culture [9]. Leaving aside the question of what is best, it is important to recognise the differences especially as big physics projects increasingly involve astronomy. Table 2, extracted from discussion at the "Great Surveys" meeting in Santa Fe in 2008, summarises some of the differences.

**Table 2**: Physics and Astronomy Culture

| Physics | Astronomy |
| --- | --- |
| Experiments | Observatories |
| Few big questions | Diverse range of studies |
| Large teams | Individuals or small teams |
| Formal structures | Informal structures |
| Formal pre-agreed author lists | PI first author |
| All participants credited | Lack of credit for experimentalists |

## 7. Conclusions

There is an increasing need for big science facilities as research areas become more mature and without the big science facilities new discoveries will decrease and the field will die. Big international facilities add extra value because they foster networking and cross fertilization, but this is offset by the increased level of bureaucracy. Small science will still prosper in the big science era and will still play a critical role opening new areas of parameter space and providing broader educational opportunities. It is also important that little science be maintained, and not displaced by big science, in the big science era.